\documentclass[letterpaper,12pt]{article}

\usepackage{times}
\usepackage{graphics}
\usepackage{amsmath,amssymb,amsfonts}
\usepackage{booktabs}
\usepackage{multirow}
\usepackage{epsfig}

\begin{document}
\pagestyle{empty}
\begin{tabular}{l}
{\Large A Likelihood Ratio Approach for Precise Discovery}\\
{\Large of Truly Relevant Protein Markers} \\[12pt] 
Lin-Yang Cheng,~~ Bowei Xi \\[12pt]
Department of Statistics, Purdue University \\[12pt]
Corresponding to: Bowei Xi, Department of Statistics, Purdue
University, \\
West Lafayette, IN, 47907. xbw@purdue.edu \\[38pt]
\end{tabular}

\noindent
{\bf Abstract}: The process of biomarker discovery
  is typically lengthy and costly, involving the phases of
  discovery, qualification, verification, and validation before
  clinical evaluation. Being able to efficiently identify the truly
  relevant markers in discovery studies can significantly simplify the
  process. However, in discovery studies the sample size is typically
  small while the number of markers being explored is much
  larger. Hence discovery studies suffer from sparsity and high
  dimensionality issues. Currently the state-of-the-art methods either find
  too many false positives or fail to identify many truly
  relevant markers. In this paper we  develop a
  likelihood ratio-based approach and aim for accurately finding
  the truly relevant protein markers in discovery studies. Our 
  method fits especially well with discovery studies because they
  are mostly balanced design due to the fact that experiments are 
  limited and controlled. Our approach is based on the observation that
the underlying distributions of expression profiles are 
unimodal for those irrelevant plain markers. 
Our method has asymptotic chi-square null
distribution which facilitates the efficient
control of false discovery rate. We then evaluate our method using both
simulated and real experimental data. In all the experiments our
method is highly effective to discover the set of truly relevant
markers, leading to accurate biomarker identifications with high
sensitivity and low empirical 
false discovery rate. \\[12pt]


\newpage

\section{Introduction}
\label{sec:intro}

\begin{figure}[h]
\centering{\includegraphics[width=4.5in]{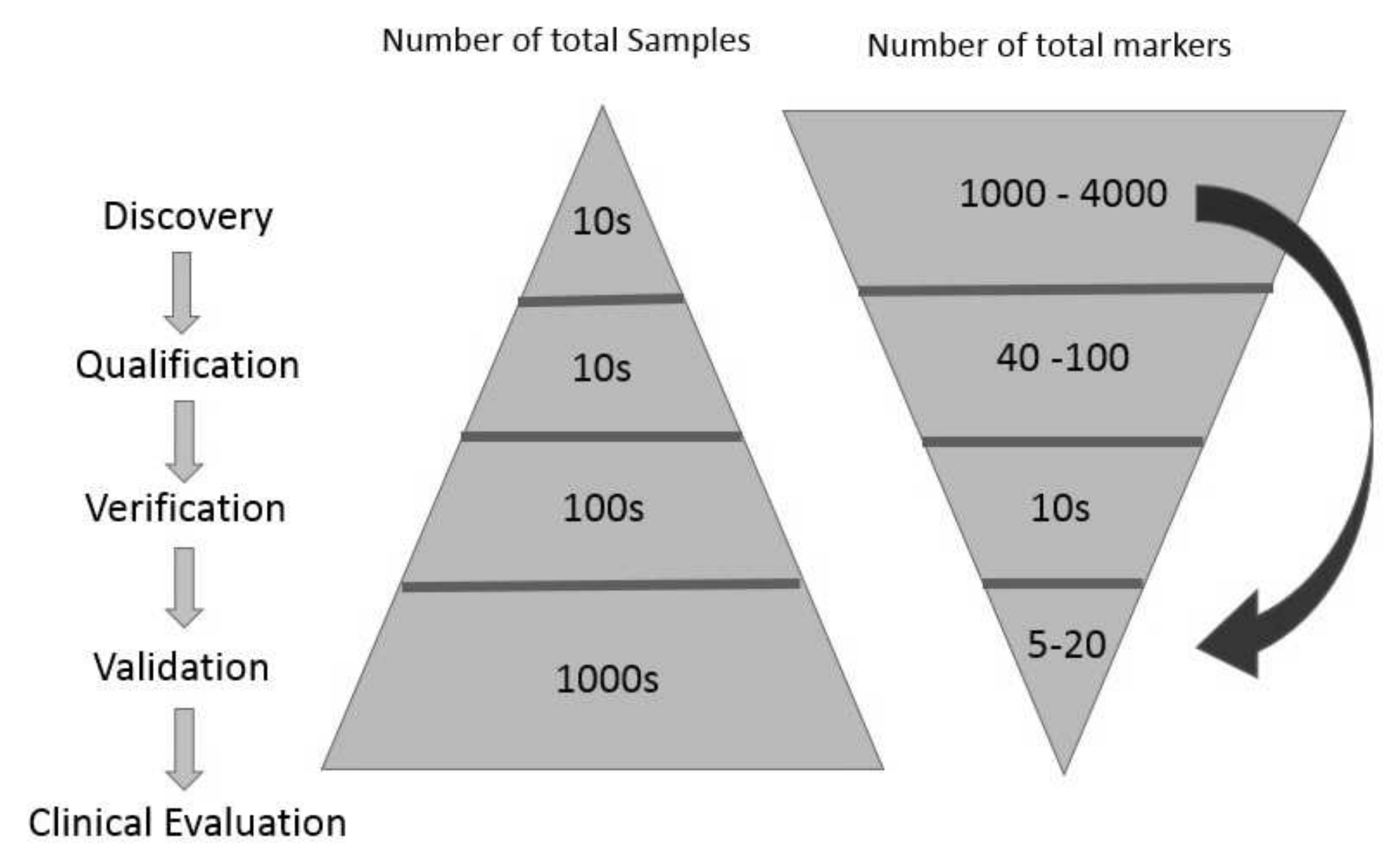}} 
\caption{The typical process of biomarker discovery.}\label{fig:Flow} 
\end{figure}

Discovering potential protein biomarkers with clinical values is
regularly a long, costly, and challenging process
\cite{Rifai2006}. Each phase of the study (i.e., discovery,
qualification, verification, and validation) costs substantial
amount of time and resources. In this paper we develop a statistical
method to expedite the process. We aim for accurately identifying
the truly relevant markers at the discovery phase, which can be
directly entered for verification or even validation phrase. 
Figure \ref{fig:Flow} illustrates our objective. 
 
Here we focus on proteomics studies. Because following a large
number of genomics studies, scientists realized that there
is not necessarily high correlation between genes and 
proteins profiles, which implies that proteomics provides
important complementary information to genomics
\cite{Pandey2000}. Moreover, since most of the drugs are
directed to interact with proteins, the protein markers are 
as informative as gene markers.  

In proteomics, mass spectrometry is considered the
most productive and reliable technique for protein
quantification. In particular, data-independent acquisition (DIA)
\cite{Carvalho2010,Gillet2012} has been proven to produce high
quality and reproducible throughput. Hence DIA is a very
effective technique for protein quantification in discovery studies. 
Currently there are a number of established univariate and
multivariate methods being applied to identify potential biomarkers
in mass spectrometry proteomics. 

LIMMA \cite{Smyth2004}, SAM
\cite{Tusher2001}, and Rank Products \cite{Breitling2004} are
popular univariate methods. LIMMA fits linear models to each
marker, then applies empirical Bayes method to form moderated
t-statistics. The moderated t-statistics are used to rank and
select the markers. SAM assigns a score to each marker based its
``relative difference''. Markers with scores larger than a
threshold value is selected. Rank Products is a non-parametric
approach, based on ranks of fold changes. Markers with the
smallest values are selected.  
Meanwhile Principal Component Analysis (PCA), Penalized Logistic
Regression (PLR), 
Partial Least Square-Discriminant Analysis (PLS-DA), and Support
Vector Machine (SVM) are well-known multivariate methods regularly used
in this field as well.    

In discovery studies the sample size is small while the
number of markers explored is large, owing to the high 
throughput of DIA.
Nowadays there are typically
thousands of protein markers quantified by DIA experiments, out
of which usually only 10-50 markers are truly relevant. 
Discovery studies encounter sparsity and high dimensionality
issues. The above univariate methods are not effective 
since they often identify too many false positives due to
sparsity issue, leading to an unacceptable high empirical false discovery 
rate (FDR). For example this is observed in  
multiple testing procedure. On the other hand the multivariate
methods mentioned above are not effective either 
because of high dimensionality issue. They often identify too few of the
truly relevant markers, leading to an unacceptable low
sensitivity. The multivariate methods mentioned above have
powerful model structures. Because they only need a small number
of markers to achieve high classification accuracy, 
these methods stop identifying more relevant markers once they
reach sufficiently accurate 
classification results. High FDR results in considerable waste of
time and resource at the subsequent phases, while low sensitivity 
leads to many unidentified useful biomarkers. Therefore, there
is a genuine need of a novel statistical method capable of 
accurately identifying the correct set of truly relevant protein
markers at the discovery phase. 

Here we develop a likelihood ratio based approach. In this paper,
we consider the scenario where are two conditions, such as
disease vs. normal. Since experiments at
the initial discovery phase are mostly controlled, our
method focuses on data from balanced design given two
conditions. Two discriminant statistics are developed based on
the likelihood ratio idea. One is for the sample size greater or
equal to five per condition. The other one is for very small
sample size, less than 5 per condition. The main idea behind our
likelihood ratio approach is that the expression profiles for the
irrelevant plain markers follow unimodal distributions, where
those for the truly relevant markers follow bimodal
distributions. The discriminant statistics are the likelihood ratios of a
bimodal distribution versus a unimodal distribution, calculated
for every protein. For larger sample size, the bimodal
distribution is approximated by a mixture of normals, and the
unimodal distribution is approximated by a univariate normal. For
very small sample size, we use kernel method to estimate the
distribution of the expression profiles, then compute the ratio
of the kernel estimate versus a univariate normal. 
Our likelihood ratio statistics have large values for
the truly relevant markers, and small values for the plain markers.  
Our proposed method is robust to deviations from
normality as long as the null expression profiles are unimodal,
even though the likelihood of normal distribution is used.
Notice with the multiple t-tests,  the null and
alternative hypotheses compare the means of two conditions. If
the null hypothesis is rejected, the two conditions have
different mean values. Essentially this indicates that marker
does not follow a unimodal distribution. Our approach goes beyond
simply considering the mean values. We explicitly estimate or
approximate the likelihood functions under the null and alternative
hypotheses. Hence our approach achieves a
much more accurate result. 

We compare our method with the aforementioned univariate and
multivariate methods for mass spectrometry-based
proteomics studies. Evaluations are performed with experimental datasets where
truly relevant markers are known from controlled mixtures or
previous literature, as well as simulated datasets.
The paper is organized as follows. 
Section~\ref{sec:method} describe our likelihood ratio
based approach for precise protein biomarker discovery. Section
\ref{sec:exp-results} shows the experimental results for our
approach compared with other
methods. Section~\ref{sec:conclusion} concludes the paper.  


%
\section{A Likelihood Ratio Approach}
\label{sec:method}

\begin{figure}[h]
\centering{\includegraphics[width=4in]{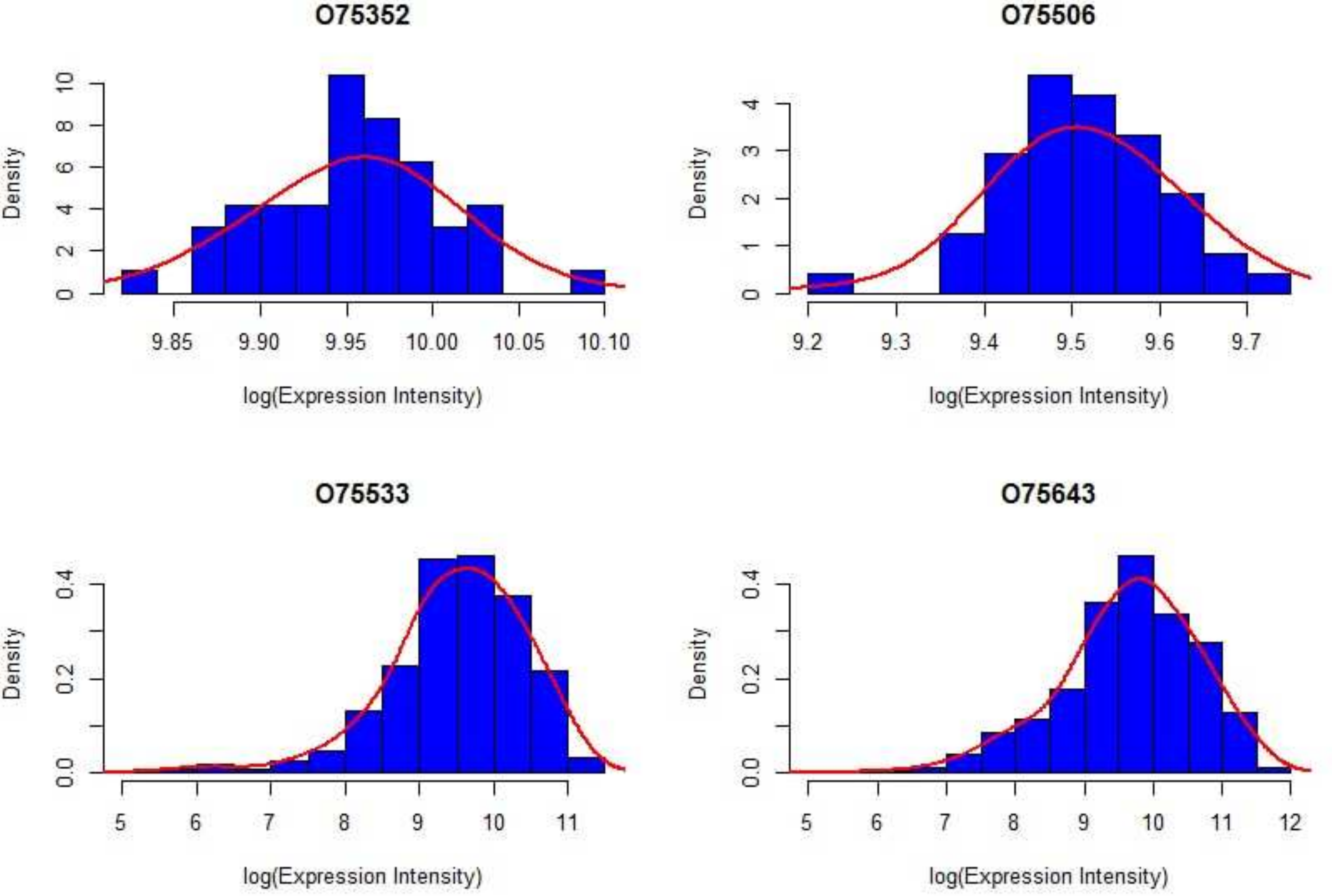}}
\caption{The expression
  distributions for four plain markers in HRM data.}\label{fig:Plain_Markers} 
\end{figure}

\begin{figure}[h]
\centering{\includegraphics[width=4in]{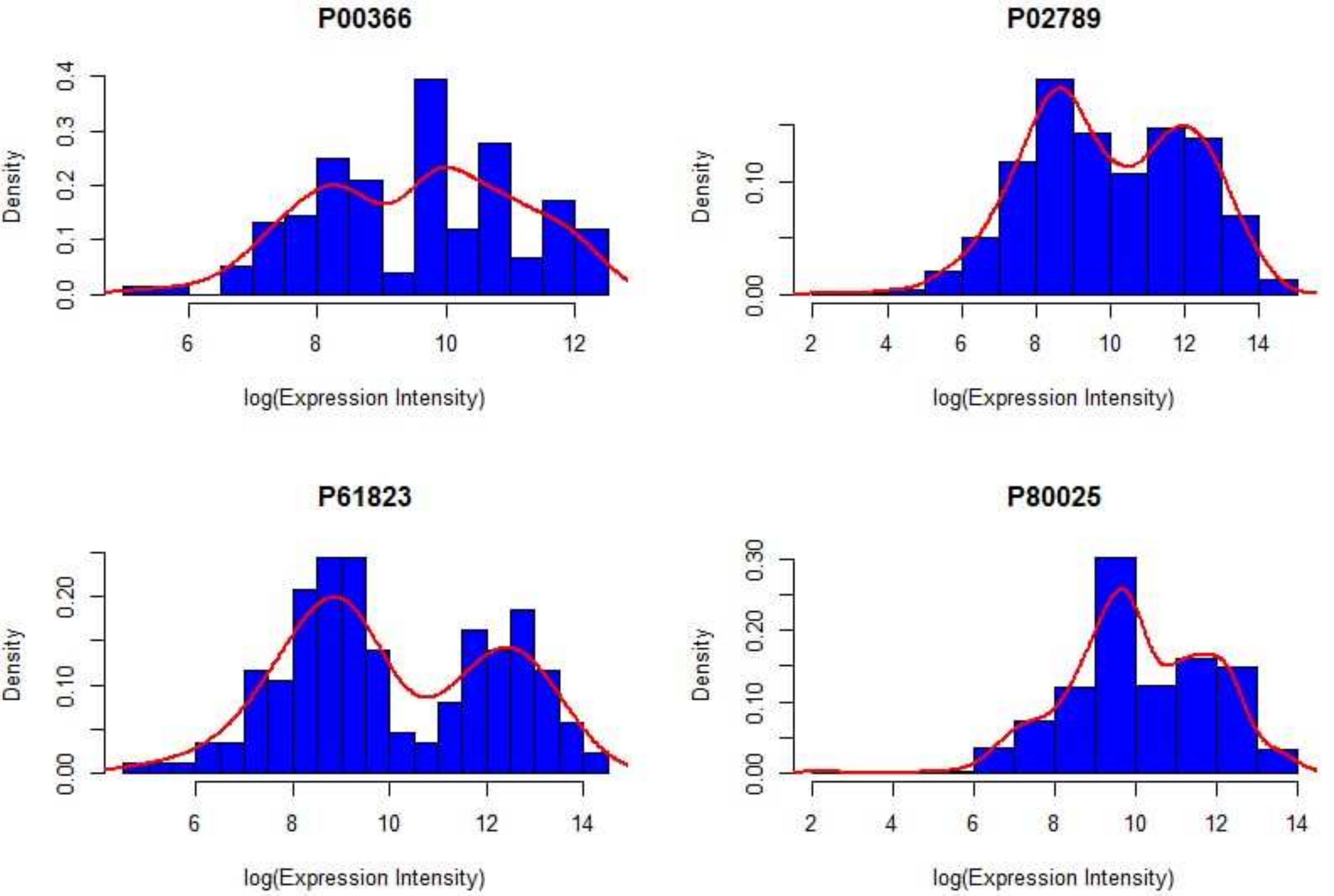}}
\caption{The expression distributions for four true 
  markers in HRM data.}\label{fig:True_Markers} 
\end{figure}

In this paper we develop two discriminant statistics which accurately
differentiate the true markers from the 
plain markers. We first examine the 
distributions of expression profiles of plain markers, as
shown in Figure \ref{fig:Plain_Markers}. These are proteins
from HRM data, discussed in Section~\ref{sec:exp-results}. Plain
protein markers from other experimental 
datasets generally share this property, i.e., their
expression profiles follow 
unimodal distributions. 
We also examine the 
distributions of expression profiles of the 
truly relevant markers, as shown in Figure
\ref{fig:True_Markers}. These are also proteins
from HRM data. True protein markers from other experimental
datasets generally follow a bimodal distribution too, given there are
two conditions.     

We discover that the distributions of expression profiles are
usually unimodal for plain markers, and bimodal for truly
relevant markers. In addition, we observe that the distributions
of expression profile for truly 
relevant markers can be approximated by a mixtures of normals, 
whereas for plain markers it can be approximated by a univariate
normal, given enough sample size to estimate the normal
parameters. Without enough sample size, we apply kernel method to
estimate the bimodal density.  

\subsection{A Discriminant Statistic for $n_i \geq 5$}
\label{sec:large-stat}
Let $n_1$ and $n_2$ be the sample size for each
condition, and $N=n_1+n_2$ be the total sample size. The mixture
proportions of the two conditions are the sample proportions,
$n_i/N$. We develop a likelihood ratio based discriminant
statistic for larger sample
size, where $n_i \geq 5$, $i=1,2$. It is
the ratio of a mixture of normals' density versus a univariate
normal density. With relatively larger sample sizes, we
approximate the bimodal distribution using a mixture of normals,
and the unimodal distribution using a univariate normal. The
likelihood ratio clearly differentiates the true markers from the
plain markers. 
  
\begin{table}[h]
\centering
\caption{Data Layout and Notation for Condition $i$, $i=1,2$}
\label{table:Design_Matrix}
\begin{tabular}{c|c|c|c}
 & Protein 1        & ...             & Protein p         \\ \hline
Sample 1      & $x^{i}_{11}$  & ...           & $x^{i}_{1p}$    \\
 $\vdots$     & $\vdots$        & $\vdots$  & $\vdots$          \\
Sample $n_i$  & $x^{i}_{n_1, 1}$  & ...    & $x^{i}_{n_1, p}$     \\ \hline          
\end{tabular}
\end{table}

Table ~\ref{table:Design_Matrix} shows a sample data layout, with
$i=1,2$. Let $\mu_1$ and $\mu_2$
be the means of each normal 
distribution composing the mixture while $\sigma^2_1$ and
$\sigma^2_2$ are their respective variances. The mixing
coefficients are $\frac{n_1}{N}$ and $\frac{n_2}{N}$.
Let $\mu$ and $\sigma^2$ be the mean and variance of the pooled
univariate normal. The discriminant statistic is calculated for
every protein. Assume there are $p$ proteins. The discriminant
statistic $R_j$ for the $j-$th protein, $j=1,...,p$, is 
calculated as follows.     
\begin{equation}
\begin{split}
R_j &=\frac{\sup\{L(\text{mixture of normals})\}}{\sup\{L(\text{univariate normal})\}}\\[10pt]
&=\frac{\sup\big\{ L\big[ \mu_1,\mu_2,\sigma^2_1,\sigma^2_2|n_1,x^{1}_{1j},...,x^{1}_{n_1, j},n_2,x^{2}_{1j},...,x^{2}_{n_2, j}\big]\big\}}{\sup\big\{L\big[ \mu,\sigma^2|N,x^{1}_{1j},...,x^{1}_{n_1, j},x^{2}_{1j},...,x^{2}_{n_2, j}\big]\big\} }\\
\end{split}
\label{eq:Discriminant_Ratio}
\end{equation}  
  
Below is the likelihood of a mixture of normals. 
\begin{equation}
\begin{split}
&L\big[\mu_1,\mu_2,\sigma^2_1,\sigma^2_2|n_1,x^{1}_{1j},...,x^{1}_{n_1,j}, 
n_2,x^{2}_{1j},...,x^{2}_{n_2, j}\big]\\[10pt]
&=\prod_{i=1}^2\big(\prod_{k=1}^{n_i}
\big(\sum_{l=1}^{2}\frac{n_l}{N\sqrt{2\pi\sigma_l^2}} 
exp\{\frac{(x^i_{kj}-\mu_l)^2}{2\sigma_l^2}\}\big)\big)\\[10pt]
\end{split}
\label{eq:Mixture_Likelihood_Individual}
\end{equation}

Below is the likelihood of the pooled normal.  
\begin{equation}
\begin{split}
&L\big[ \mu,\sigma^2|N,x^{1}_{1j},...,x^{1}_{n_1, j},x^{2}_{1j},...,x^{2}_{n_2, j}\big]\\[10pt]
&=\prod_{i=1}^2\prod_{k=1}^{n_i}
\big(\frac{1}{\sqrt{2\pi\sigma^2}} 
exp\{\frac{(x^i_{kj}-\mu)^2}{2\sigma^2}\}\big)\\[10pt]
\end{split}
\label{eq:Univariate_Likelihood}
\end{equation}


The maximum of the likelihood function is computed by plugging in
the maximum likelihood estimates (MLE) of parameters to the
likelihood function. Let $\hat{\mu}_1$, $\hat{\mu}_2$,
$\hat{\sigma}^2_1$, and $\hat{\sigma}^2_2$ be the MLEs. MLEs are
obtained by maximizing the corresponding likelihood functions. 
The MLEs have closed form \cite{Bilmes1998} as follows. We have
the MLEs for the pooled normal as follows. 
$$
\hat{\mu} = \frac{\sum_{i,k,j} x^{i}_{kj}}{N},
$$
$$
\hat{\sigma^2} = \frac{\sum_{i,k,j} (x^{i}_{kj}-\hat{\mu})^2}{N}.
$$
Let $f(x^{i}_{kj},\mu_l,\sigma^2_l)=  
exp\{\frac{(x^i_{kj}-\mu_l)^2}{2\sigma_l^2}\}/\sqrt{2\pi\sigma_l^2}$. $l=1,2$. Let
$$
p_l(x^{i}_{kj}) =
\frac{f(x^{i}_{kj},\mu_l,\sigma^2_l)}{\sum_{l=1}^2 (f(x^{i}_{kj},\mu_l,\sigma^2_l)n_l)}. 
$$
The MLEs for the mixture of normals given the mixing coefficients
$\frac{n_1}{N}$ and $\frac{n_2}{N}$ are the following, $l=1,2$. 
$$
\hat{\mu}_l = \frac{\sum_{i,k,j}(p_l(x^{i}_{kj})\times x^{i}_{kj}) }{\sum_{i,k,j} p_l(x^{i}_{kj})},
$$
$$
\hat{\sigma}^2_l = \frac{\sum_{i,k,j} (p_l(x^{i}_{kj})\times (x^{i}_{kj}-\hat{\mu}_l)^2) }{\sum_{i,k,j} p_l(x^{i}_{kj})}
$$
The discriminant statistic, $R_j$, for the $j-$th protein is the following. 
\begin{equation}
\begin{split}
R_j &=\frac{L\big[ \hat{\mu}_1,\hat{\mu}_2,\hat{\sigma}^2_1,\hat{\sigma}^2_2|n_1,x^{1}_{1j},...,x^{1}_{n_1, j},n_2,x^{2}_{1j},...,x^{2}_{n_2, j}\big]}{L\big[ \hat{\mu},\hat{\sigma}^2|N,x^{1}_{1j},...,x^{1}_{n_1, j},x^{2}_{1j},...,x^{2}_{n_2, j}\big] }\\[10pt]
\end{split}
\label{eq:Discriminant_Statistic}
\end{equation}  
Consequently, from \eqref{eq:Discriminant_Statistic} we can tell
that $R_j$ is expected to have a large value if protein j is a truly
relevant marker. Otherwise, $R_j$ is expected to be small. 

\subsection{Asymptotic Property of $R_j$}
Here we derive the asymptotic property of the statistic $R_j$
under the null hypothesis, which enables us to 
quantify the statistical significance on each
protein provided sample size is sufficient.  
The discriminant statistic $R_j$ can be thought of as the
statistic for testing whether the $j-$th protein truly has expression
profiles correlate with conditions or not, with the following
null and alternative hypotheses.

\begin{equation*}
\begin{split}
\begin{cases}
H_0: \ &\text{The $j-$th protein is a plain marker}\\ 
 & \text{(unimodal
  approximated by univariate
  normal)} \\
H_a: \ &\text{The $j-$th protein is a relevant marker}\\ 
 & \text{(bimodal
  approximated by a mixture of normals) }
\end{cases}
\end{split}
\end{equation*}  
Hence we write the 
hypotheses in terms of notations as follows.
\begin{equation}
\begin{split}
\begin{cases}
H_0: \ &\mu_1=\mu_2~~~and~~~\sigma^2_1=\sigma^2_2\\
H_a: \ &\mu_1\neq\mu_2~~~or~~~\sigma^2_1\neq\sigma^2_2
\end{cases}
\end{split}
\label{eq:Hypothesis_Marker_notation}
\end{equation}  
The statistic for the $j-$th protein 
$R_j=\sup[L(H_a)]/\sup[L(H_0)]$ is specified in
Equation~\ref{eq:Discriminant_Statistic}. 
We note that $R_j$ is the
likelihood ratio statistic for testing between full and
reduced models. Therefore, the quantity $2\ln (R_j)$ is the
deviance of the goodness-of-fit test. As the asymptotic property
of the deviance had been known, we know that the asymptotic null
distribution of $2\ln(R_j)$ is chi-square.  
Note that the degrees of freedom is the difference in dimension
of parameter spaces between $H_a$ and $H_0$. In particular, the
statistic $2\ln (R_j)$ has 2 degrees of
freedom because its corresponding hypotheses in Equation 
\eqref{eq:Hypothesis_Marker_notation} has 4 free parameters in
the distribution under $H_a$ and only 2 in the distribution under
$H_0$. We have  
\begin{equation}
\begin{split}
2\ln (R_j)=2\ln\big\{\frac{\sup[L(H_a)]}{\sup[L(H_0)]}\big\} \xrightarrow{d} \chi^2_2 \quad \text{under} \ H_0
\end{split}
\label{eq:R_Aymptotic}
\end{equation}  

Although the property of $2\ln(R_j)$ is asymptotic in theory, in
practice it takes around 10 to 30 samples per condition to converge to
chi-square, assuming a balanced design ($n_1=n_2=n$) as it is the most common
case in discovery studies.  Table
\ref{table:Rj_Convergence} shows the empirical tail
probability $P[R_j>\chi^2_{2,\alpha}]$, with
$\alpha=0.01,0.05,0.1$, using simulated data discussed in 
Section~\ref{Simulation}. It demonstrates that the convergence of $2\ln(R_j)$
to $\chi^2_2$ is valid and efficient. In general, this
convergence is satisfactory when there are at least 10 samples
per condition (n$\geq$10).  

\begin{table}[h]
\centering
\caption{The empirical tail probability of $R_j$ with increasing
  sample sizes. This verifies the 
  asymptotic property of $R_j$ in \eqref{eq:R_Aymptotic} and
  shows the fast convergence rate.} 
\label{table:Rj_Convergence}
\begin{tabular}{c|c|c|c}
$n_i$  & $P[R_j>\chi^2_{2,0.01}]$  & $P[R_j>\chi^2_{2,0.05}]$ & $P[R_j>\chi^2_{2,0.10}]$ \\ \hline
3            & 0.22                     & 0.30                     & 0.33 \\
5            & 0.10                     & 0.15                     & 0.21 \\
8            & 0.04                     & 0.09                     & 0.16 \\
10           & 0.03                     & 0.08                     & 0.13 \\
15           & 0.02                     & 0.07                     & 0.11 \\
20           & 0.02                     & 0.05                     & 0.10 \\
30           & 0.01                     & 0.05                     & 0.10
\end{tabular}
\end{table}

\subsection{Alternative Statistic for Very Small Sample Size $n_i<5$}
When there are at least five samples per condition, the sample means
and the sample variances are sufficiently accurate estimates. Consequently
we obtain a reasonable likelihood function of a mixture normal to
be used in the discriminant statistic in
Section~\ref{sec:large-stat}. However discovery studies usually face 
small sample sizes. Sometimes there are less than five sample per
condition, which leads to inaccurate sample means and sample
variances. When the sample size is too small, the statistic
in Section~\ref{sec:large-stat} does not perform well. We notice
that plain markers have unimodal expression file
distributions is a strong identifying feature. Here we
develop an alternative statistic which performs well for very
small sample size, with $n_1=n_2=n<5$. 
The idea for very small sample size is to compare the estimated empirical
distribution of a protein's expression profile with a pooled
univariate normal, where the samples from two conditions are combined. The
null hypothesis is that the particular protein in question is
plain. We use kernel density estimation method for approximating the
empirical distribution of protein expression profiles, which
provide results quite sensitive to non-unimodal patterns. Notice
for larger sample size, the estimated empirical distribution can be
simplified to an approximate mixture of normals.  

For the $j-$th protein, we estimate its expression profile
distribution's density function using the normal kernel as follows.
\begin{equation}
\begin{split}
\hat{g}_{h_j}(x)=\frac{1}{nh_j}\sum_{i=1}^2\sum_{k=1}^{n} K(\frac{x-x^{i}_{kj}}{h_j}), 
\end{split}
\label{eq:Kernel_Density}
\end{equation}  
where that $n_1=n_2=n$ is the sample size of each condition;
$x^{i}_{kj}$ is the expression intensity as displayed in
Table~\ref{table:Design_Matrix}; K($\cdot$) is the density of
standard normal; and $h_j$ is the optimal bandwidth for the
$j-$th protein derived according to
minimum mean integrated squared error as follows. 
\begin{equation}
\begin{split}
h_j=\operatorname*{arg\,max}_{h_j} E\{\int [\hat{g}_{h_j}(x)-g_j(x)]^2 dx \}, 
\end{split}
\label{eq:Kernel_Bandwidth}
\end{equation}  
where $\hat{g}_{h_j}(x)$ is the kernel density estimation from
\eqref{eq:Kernel_Density} with bandwidth of $h$ whilst $g_j(x)$ is
the (unknown) true density of expression profile. 

As opposed to simply testing the parameters when the sample size
is at least five per condition, as shown in
\eqref{eq:Hypothesis_Marker_notation}, here we
compare two distributions. For the $j-$th protein
density function $g_{h_j}(x)$, let $G_{h_j}(x)$ be the
corresponding cumulative distribution function (CDF). Under the
alternative hypothesis $G_{h_j}(x)$ does not follow a unimodal
distribution. Let $F_j(x)$ be the CDF of the pooled univariate normal under the
null hypothesis that the observed expressions are drawn from
a single univariate normal distribution. The null and the
alternative hypothesis are written as follows.
\begin{equation*}
\begin{split}
\begin{cases}
H_0: \ G_{h_j}(x)=F_j(x) \ \ (\text{The $j-$th protein is a plain marker})\\
H_a: \ G_{h_j}(x)\neq F_j(x) \ \ (\text{The $j-$th protein is
  truly relevant})
\end{cases}
\end{split}
\label{eq:Hypothesis_Marker_SmallSample}
\end{equation*}   
We then use the Kolmogorov-Smirnov (K-S) method for testing
the hypotheses. K-S
test uses the maximum difference between the CDF of kernel
density and that of univariate normal, both evaluated at each
observed expression, as the test statistic. 
\begin{equation}
\begin{split}
D_j=\operatorname*{max}_{i, k} \lvert G_{h_j}(x^{i}_{kj})-F(x^{i}_{kj})   \rvert
\end{split}
\label{eq:KS_Test_Statistic}
\end{equation} 

Massey~\cite{Massey1951} claimed that K-S test is superior to the
chi-square test especially in cases of small sample size. Since
the normal CDF $F_j(x)$ in hypotheses
\eqref{eq:Hypothesis_Marker_SmallSample} is not specified in
advance (i.e., the mean and variance are estimated from the
data), the significance level and p-value of K-S test should be
derived carefully. Hence, we apply the method developed in 
Lilliefors~\cite{Lilliefors1967} to obtain the p-value as
mean and variance of univariate normal are not 
specified in advance but estimated from the sample. 

\section{Experimental Results}
\label{sec:exp-results}

\subsection{Experimental Datasets}
\label{sec:data}

The following two experimental datasets are used to evaluate the
performance of our approach in
Section~\ref{sec:exp-results}. In this section, we use two
criteria to compare different approaches. One is sensitivity,
which is the ratio of the number of truly relevant markers among the
selected markers vs the actual number of truly relevant
markers. The other is FDR, which is the ratio of 
false positives among the selected markers vs the number of
selected markers.  
The truly relevant protein markers
are known in the two datasets. 
Hence we are able to compare the
sensitivity and FDR of different approaches including ours. 
 
\paragraph{\bf{Hyper Reaction Monitoring (HRM) Data:}}
Hyper Reaction Monitoring (HRM) is a novel DIA technique proposed
by \cite{Bruderer2015} where spectral libraries were based on
normalized retention time for data extraction. The main
contribution of \cite{Bruderer2015} 
is to show the capability of their DIA technique.
Compared with DDA technique, they demonstrated the increased
capability that DIA technique brings to protein biomarker discovery.
In order to show 
that HRM is superior to the established old-generation DDA
technique, they conducted experiment where
3785 proteins were quantified and 12 of them were known to be
correlated with conditions.  
This is a typical setup for discovery study in that there are 8
conditions with 3 samples in each condition. Later in our
experiments,
we merge the samples from similar conditions so that there are 2
conditions with 12 samples each.  

\paragraph{\bf{Yeast Data:}}
SWATH-MS \cite{Selevsek2015} was invented as an innovative DIA
experiment, which combined the strength of DIA and target data
analysis for more throughput with high accuracy. In the interest of
proving the competence of SWATH-MS they quantified protein
markers on \textit{Saccharomyces cerevisiae} with osmotic stress
imposed over time. There are 6 time points, regarded as
conditions, and 3 samples in each condition. In total, 2602
proteins were quantified in which 50 were known to be regulated
in response to osmotic stress. In addition, proteins responding
to osmatic stress are mainly associated with arbohydrate and
amino acid metabolism. Therefore, the ability SWATH-MS provides to recover known
biomarkers shows that DIA data has promising potential to assist
in discovering relevant protein markers. 

This experimental data
is another typical discovery study, where many markers are
explored while sample size is limited. In our experiments, we
again merge the samples from the first 3 time points, and those
from the last 3 time points. We then have 2
condition where there are 9 samples each. Merging the samples is
reasonable here since the change of protein expressions, if any,
happens gradually as the experiment progressed. 

\begin{figure}[h]
\centering{\includegraphics[width=3in]{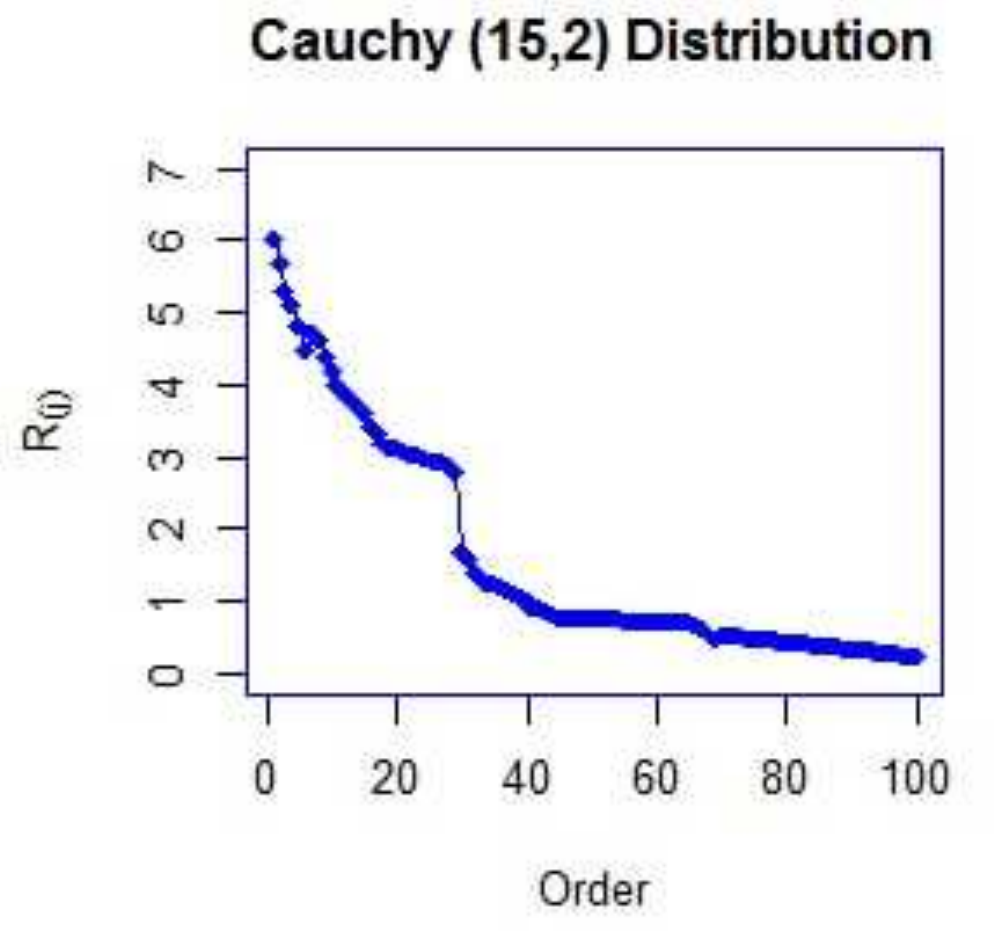}}
\caption{Ordered values of the statistic $R_j$. The largest 100
  are shown.} 
\label{fig:R_j_Order_Cauchy}
\end{figure}

\begin{figure}[h]
\centering{\includegraphics[width=3in]{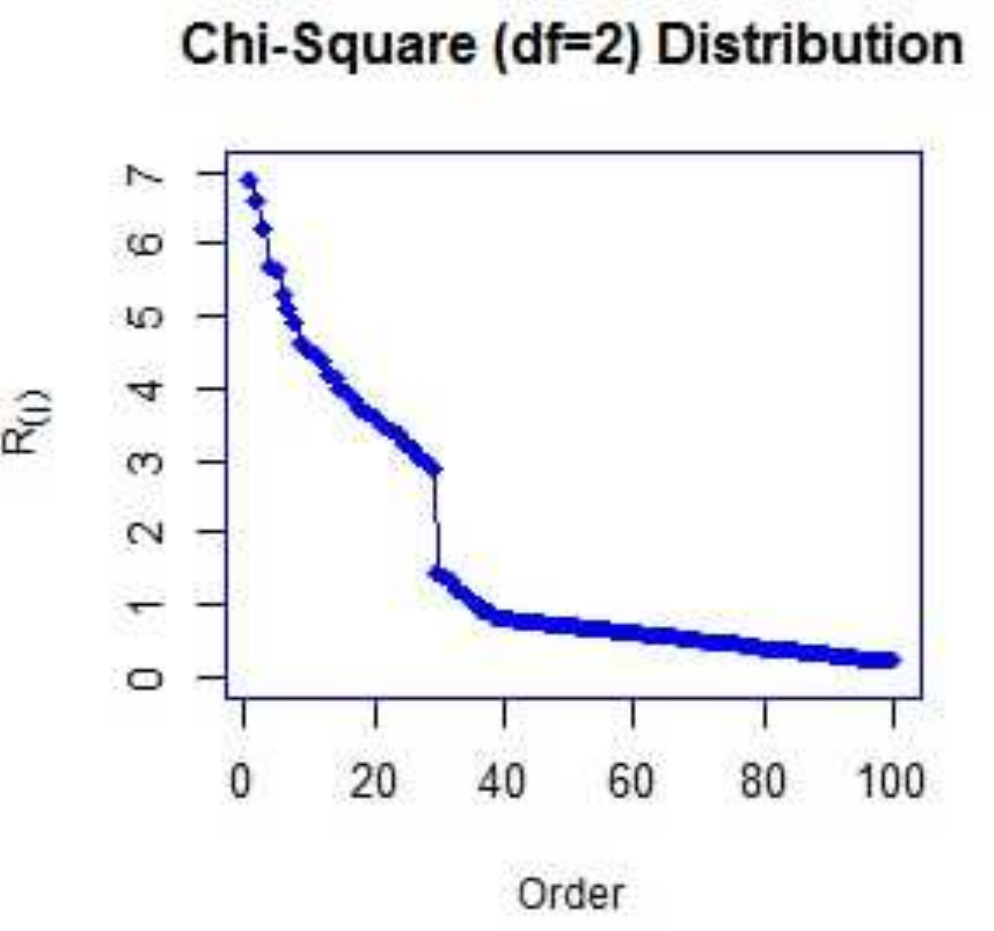}}
\caption{Ordered values of the statistic $R_j$. The largest 100
  are shown.} 
\label{fig:R_j_Order_ChiSquare}
\end{figure}

\subsection{Simulated Datasets} 
\label{Simulation}
We also simulate a large number of datasets with different
configurations to 
evaluate our approach. In general, in each simulation run 
1000 protein markers are generated, and the expression profiles
of each protein are simulated according to the following steps. Note that
the parameters we use are the estimated values of those from the data
sets acquired and processed by OpenSWATH \cite{Rost2014}, where
proteins were under the matrices of water and
human. Water is the simplest matrix among them, with clean
signals, whereas human is the most complex under 
considerable interference. Simulated data points are denoted as
$X^{i}_{kj}$, which is the expression intensity for the $j-$th
protein of the $k-$th subject within the $i-$th condition. 
An example of simulated data layout is shown in Table
~\ref{table:Design_Matrix}. 

\begin{enumerate}

\item{Step 1.} Generate the expression intensity $X^{i}_{kj}$
  according to the following equation~(\cite{Rost2014}). 
$$ X^{i}_{kj}=\mu+C_i+S(C)_{k(i)}+F_j+\epsilon^{i}_{kj} $$

\item{Step 2.} In the equation above $C_i$ stands for the main effect of a condition,
$S(C)_{k(i)}$ the effect of each subject within a condition, and
  $F_j$ the main effect of
each protein. We then generate the source of variations for the
true markers and plain markers as follows. We simulate 2 
conditions, $i=1,2$. 

For true markers $C_1\neq C_2$. For plain markers $C_1=C_2$. 
Then the other items in the above equation are simulated for
normal distributions as follows. 
$$S(C)_{k(i)}\sim N(0, \sigma^2_S),$$
$$F_j\sim N(0, \sigma^2_F),$$
$$\epsilon^{i}_{kj}\sim N(0, \sigma^2).$$

\item{Step 3.} Set the parameter values according to~\cite{Rost2014}. We
  have $\mu=15$, $\sigma^2_S=27.37$, and $\sigma^2_F=0.98$. The sample
  size for each condition and the number of proteins vary
  in different simulation runs. The error term $\epsilon^{i}_{kj}$'s
  variance depends on the matrices. We have $\sigma^2=2.23$ for
  human background, the most complex matrix; $\sigma^2=0.48$ for
  water background, the matrix 
  with the smallest level of interference.    

\end{enumerate}

\begin{figure}[h]
\centering{\includegraphics[width=3in]{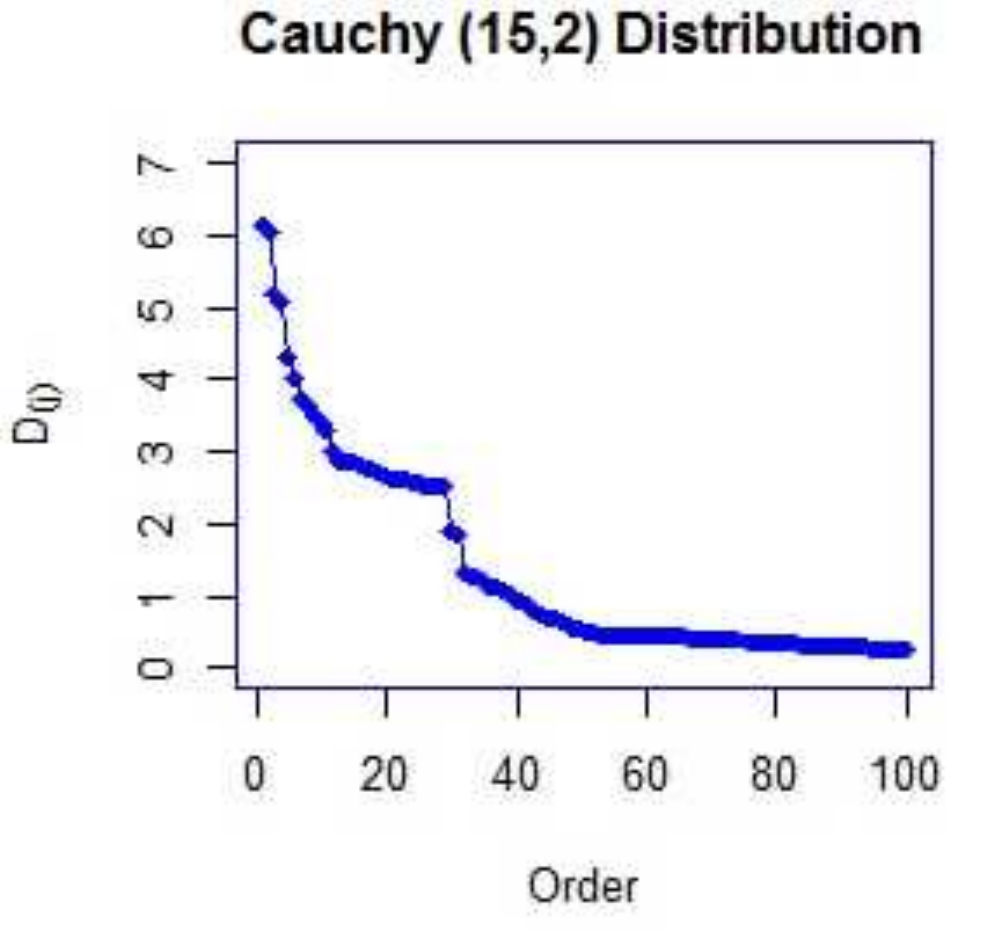}}
\caption{Ordered values of the statistic $D_j$. The largest 100
  are shown.} 
\label{fig:D_j_Order_Cauchy}
\end{figure}

\begin{figure}[h]
\centering{\includegraphics[width=3in]{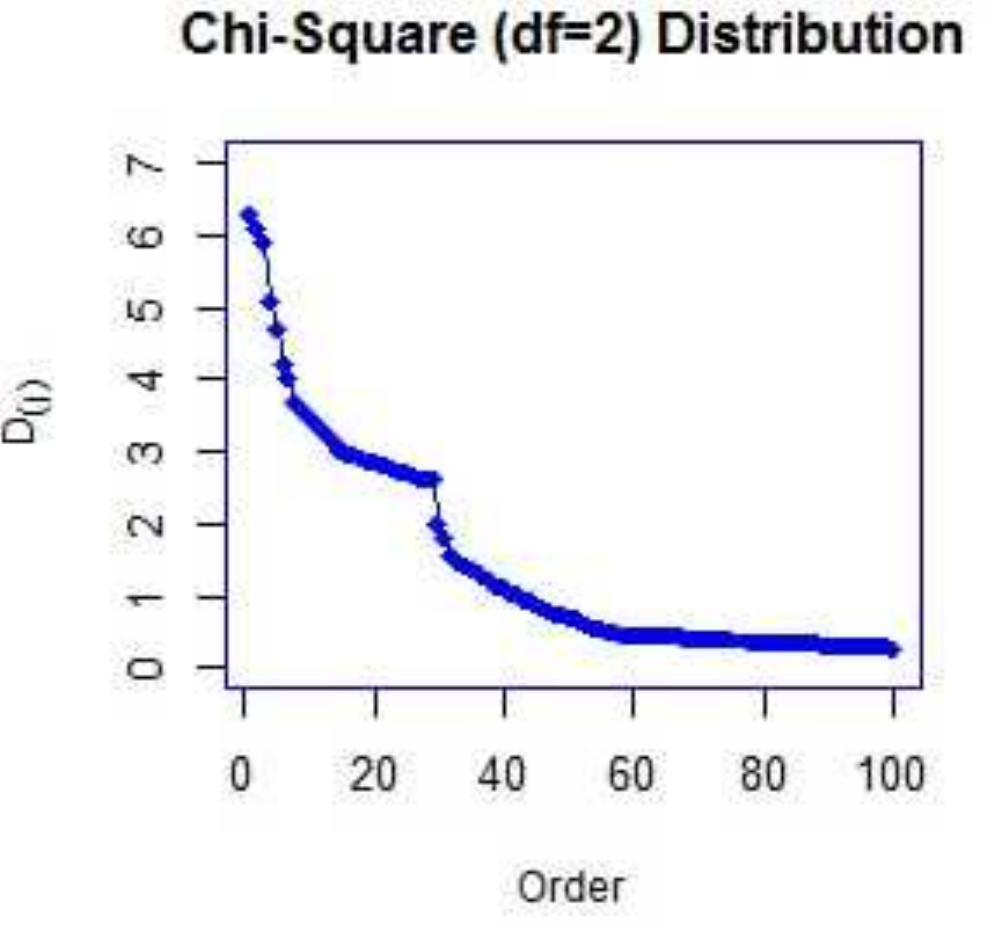}}
\caption{Ordered values of the statistic $D_j$. The largest 100
  are shown.} 
\label{fig:D_j_Order_ChiSquare}
\end{figure}

%

\subsection{Robust Performance Against Deviations from Normality}
Although the two statistics in Section~\ref{sec:large-stat} are 
constructed as a ratio of a 
mixture distribution or the empirical distribution versus a
pooled univariate normal, they show 
robust performance with non-normal distributions too. 
Figure~\ref{fig:Plain_Markers} shows the  
empirical expression profile distributions of plain markers in
\textit{HRM data}. The most important feature of plain markers is
that their expression profile distributions are unimodal. Hence
the fundamental idea behind our approach is to separate 
bimodal distributions from unimodal distributions, which lead
to accurate identification of true markers. The underlying actual 
distributions do not need to strictly follow normality assumptions for our
approach to perform well. Here we 
demonstrate the robust performance of the two statistics against
strong deviations from normality. 

In one experiment, we have 10 samples per
condition. We simulate 1000 proteins as specified in
Section~\ref{Simulation} but with different error term
distributions, where there are 30 true markers and 970 
plain markers. The error terms are
simulated according to Cauchy distribution with location
parameter 15 and scale parameter 2 (Figure~\ref{fig:R_j_Order_Cauchy})
and chi-square distribution 
with two degrees of freedom
(Figure~\ref{fig:R_j_Order_ChiSquare}). Cauchy distribution has  
much heavier tails than normal and chi-square with two degrees
of freedom is very skewed distribution.  
Plain markers have $C_1=C_2=U$ where $U$ is randomly sampled following
a uniform distribution over $[1,100]$. 
The effect size is 3 fold for true markers. The fold change is
defined on log base 2 scale. True markers have $C_1=\log_2(6)+U$
and $C_2=\log_2(2)+U$  where $U$ is randomly sampled from 
a uniform distribution over $[1,10]$.  The
resulting expression profile distributions for both plain markers 
and true markers are strongly non-normal. 

The proteins are ordered by their $R_j$ values. The largest 100
proteins are shown in Figures~\ref{fig:R_j_Order_Cauchy} and
~\ref{fig:R_j_Order_ChiSquare}.  
The ordered values of the statistic $R_j$, which works well for
not-too-small sample size, show a sudden drop after the  
largest values. We choose the cutoff for $R_j$ at where the drop
occurs. This pattern is also observed from the two experimental
datasets in Section~\ref{sec:exp-tables}. We select the markers
with $R_j$ values larger than the cutoff. Based on this selection
method, we are able to accurately 
identify 1) the number of truly relevant markers; 2) the relevant
markers themselves. 

In another experiment we simulate 1000 proteins as in the first
experiment, only there are 3 samples per conditions. The
statistic $D_j$ for very small sample size is
used. Figures~\ref{fig:D_j_Order_Cauchy}
and~\ref{fig:D_j_Order_ChiSquare} show the order values of $D_j$,
the 100 largest ones, which exhibit the same pattern. With very small sample size, we
are still able to accurately identify the relevant markers. 


\subsection{High Sensitivity and Low FDR}
\label{sec:exp-tables}

\begin{figure}[h]
\centering{\includegraphics[width=3in]{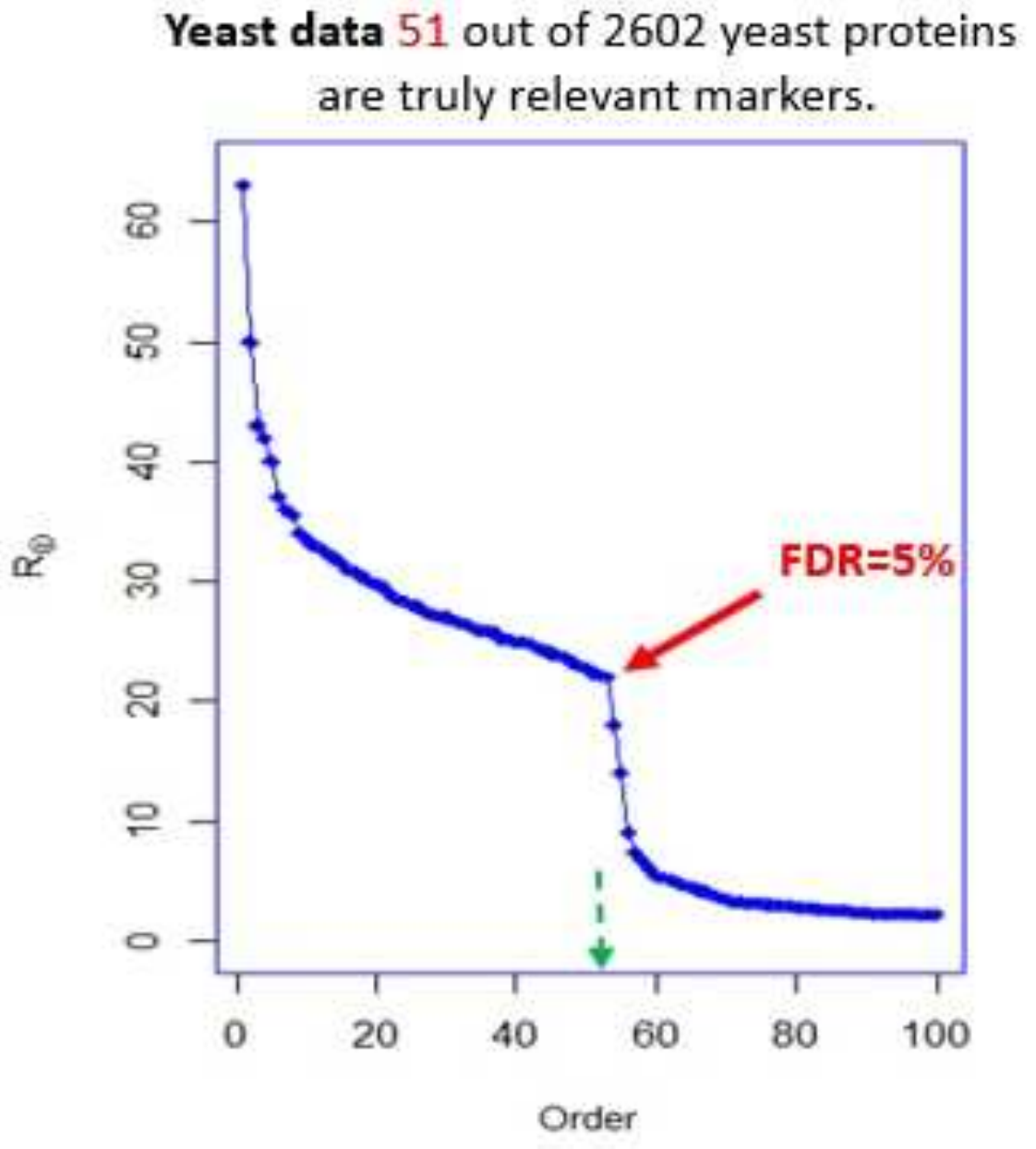}}
\caption{Ordered values of the statistic $R_j$ for Yeast Data. The largest 100
  are shown.} 
\label{fig:Rj-yeast}
\end{figure}

\begin{figure}[h]
\centering{\includegraphics[width=3in]{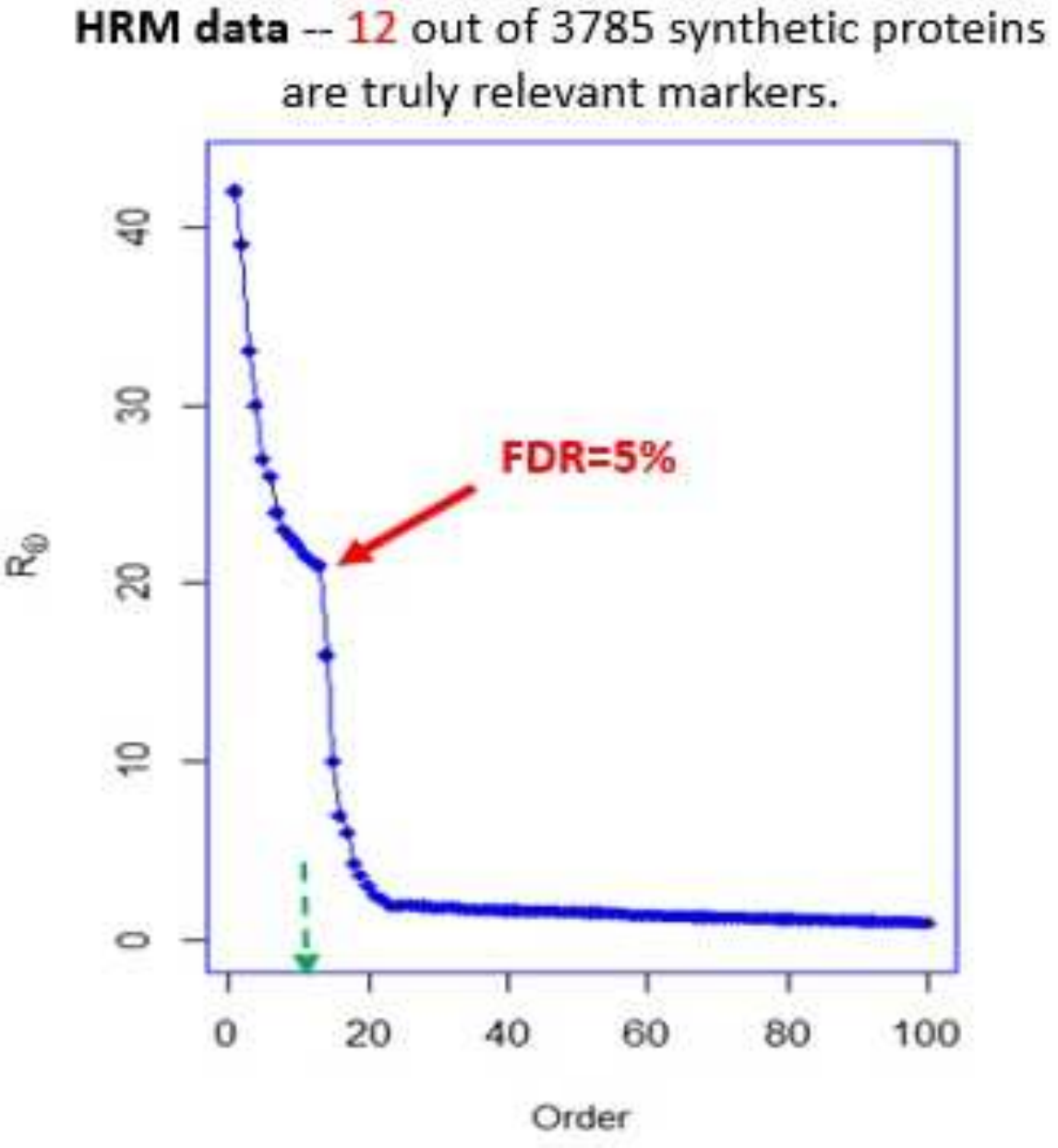}}
\caption{Ordered values of the statistic $R_j$ for HM Data. The largest 100
  are shown.} 
\label{fig:Rj-HRM}
\end{figure}

\begin{table}[h]
\centering
\caption{Sensitivity and empirical FDR for Yeast and HRM data}  
\label{table:SEN_FDR_Uni_Multi_Proposed}
\begin{tabular}{l|ll|ll}
       & \multicolumn{2}{c|}{ \textit{HRM data}} & \multicolumn{2}{c}{ \textit{Yeast data}} \\
       & Sensitivity          & Empirical FDR        & Sensitivity           & Empirical FDR             \\ \hline
LIMMA  & $\frac{12}{12}=1.00$ & $\frac{14}{26}=0.54$ & $\frac{51}{51}=1.00$  & $\frac{36}{87}=0.41$   \\
RP     & $\frac{12}{12}=1.00$ & $\frac{9}{21}=0.43$  & $\frac{48}{51}=0.94$  & $\frac{31}{82}=0.38$      \\
SAM    & $\frac{12}{12}=1.00$ & $\frac{8}{20}=0.40$  & $\frac{50}{51}=0.94$  & $\frac{41}{91}=0.45$      \\
t-test & $\frac{12}{12}=1.00$ & $\frac{16}{28}=0.57$ & $\frac{50}{51}=0.98$  & $\frac{55}{106}=0.52$     \\ \hline    
PCA    & $\frac{6}{12}=0.50$ & $\frac{1}{6}=0.17$ & $\frac{8}{51}=0.16$  & $\frac{1}{8}=0.13$   \\
PLR    & $\frac{3}{12}=0.25$ & $\frac{0}{3}=0$    & $\frac{6}{51}=0.12$  & $\frac{0}{6}=0$      \\
PLS-DA & $\frac{7}{12}=0.58$ & $\frac{0}{7}=0$    & $\frac{11}{51}=0.22$  & $\frac{1}{12}=0.08$      \\
SVM    & $\frac{7}{12}=0.58$ & $\frac{0}{8}=0$    & $\frac{15}{51}=0.29$  & $\frac{1}{16}=0.06$     \\     \hline
Our $R_j$ &$\frac{12}{12}=1.00$&$\frac{1}{13}=0.08$&$\frac{50}{51}=0.98$&$\frac{4}{54}=0.07$
\end{tabular}
\end{table} 

We apply the statistic $R_j$ to the two experimental datasets,
Yeast data and HRM data as discussed
earlier. Figures~\ref{fig:Rj-yeast} and~\ref{fig:Rj-HRM} show
the ordered values of $R_j$. Again we observe a sudden drop of
$R_j$ values. We set the cutoff
of $R_j$ to be 2.5. We then select the
proteins that have $R_j$ values larger than cutoff. 
In Table \ref{table:SEN_FDR_Uni_Multi_Proposed} we compare
the commonly used univariate and
multivariate methods with 
our statistic $R_j$ based on sensitivity and FDR. For univariate
methods FDR cutoff is 
set at 0.05, while for multivariate methods the algorithms stop
based on their default stopping criteria. 
The univariate methods select too many markers, and suffer from
high empirical FDR. The multivariate methods have strong model
structures. They do not need all the relevant markers to achieve
high classification accuracy. Hence they select too few markers,
leading to very low sensitivity. Our approach, to cut off at where $R_j$
values suddenly drop, returns superior sensitivity with very
low FDR.  
  
Next we compare our approach with the commonly used methods in
simulation studies. In a single run we simulate 1000 proteins as specified in
Section~\ref{Simulation} with the error terms normally
distributed, where there are 30 true markers and 970  
plain markers.  
Again plain markers have $C_1=C_2=U$ where $U$ is randomly sampled following
a uniform distribution over $[1,100]$. 
The effect size is 3 fold for true markers. True markers have $C_1=\log_2(6)+U$
and $C_2=\log_2(2)+U$ where $U$ is randomly sampled following
a uniform distribution over $[1,10]$. In one experiment we have
10 samples per condition and conduct 50 runs, with the average
sensitivity and the average FDR shown in
Table~\ref{table:sim-compare-10}. In another experiment we have 3
samples per condition and conduct 50 runs, with the average
sensitivity and the average FDR shown in
Table~\ref{table:sim-compare-3}. 

\begin{table}[h]
\centering
\caption{Average sensitivity and empirical FDR for water, human, yeast
  background with 10 samples per condition. FDR is set at
  0.05 for univariate methods. $R_j$ cutoff is 2.5}  
\label{table:sim-compare-10}
\begin{tabular}{l|ll|ll}
       & \multicolumn{2}{c|}{ \textit{Water}} & \multicolumn{2}{c}{ \textit{Human}} \\
       & Sensitivity          & Empirical FDR        & Sensitivity           & Empirical FDR             \\ \hline
LIMMA  & $0.87$ & $0.50$  & $0.67$  & $0.40$   \\
RP     & $0.73$ & $0.41$   & $0.80$  & $0.33$      \\
SAM    & $0.90$ & $0.44$  & $0.77$  & $0.43$      \\
t-test & $0.70$ & $0.29$   & $0.50$  & $0.40$     \\ 
\hline    
PCA    & $0.17$  & $0.20$    & $0.13$  & $0.25$   \\
PLR    & $0.40$ & $0.25$   & $0.33$  & $0.1$      \\
PLS-DA & $0.23$  & $0.29$    & $0.13$  & $0.04$      \\
SVM    & $0.40$ & $0.17$   & $0.37$  & $0.09$     \\     \hline
Our $R_j$  &$0.90$&$0.15$&$0.77$&$0.17$
\end{tabular}
\end{table} 

\begin{table}[h]
\centering
\caption{Average sensitivity and empirical FDR for water, human, yeast
  background with 3 samples per condition. FDR is set at
  0.1 for univariate methods. $D_j$ cutoff is 2.5}  
\label{table:sim-compare-3}
\begin{tabular}{l|ll|ll}
       & \multicolumn{2}{c|}{ \textit{Water}} & \multicolumn{2}{c}{ \textit{Human}} \\
       & Sensitivity          & Empirical FDR        & Sensitivity           & Empirical FDR             \\ \hline
LIMMA  & $0.40$ & $0.54$  & $0.48$  & $0.51$   \\
RP     & $0.53$ & $0.43$  & $0.55$  & $0.58$      \\
SAM    & $0.37$ & $0.40$  & $0.40$  & $0.40$      \\
t-test & $0.23$ & $0.57$  & $0.43$  & $0.41$     \\ 
\hline    
PCA    & $0.37$ & $0.37$    & $0.19$  & $0.13$   \\
PLR    & $0.30$ & $0.13$    & $0.13$  & $0.16$      \\
PLS-DA & $0.41$ & $0.09$    & $0.38$  & $0.40$      \\
SVM    & $0.50$ & $0.25$    & $0.45$  & $0.35$     \\     \hline
Our $D_j$  &$0.67$&$0.20$     &$0.53$   &$0.31$
\end{tabular}
\end{table} 

\clearpage
\section{Conclusion and Discussion}
\label{sec:conclusion}
Our approach is superior to the popularly applied
methods univariate and multivariate methods given reasonable
sample size. Our approach shows
good performance for very small sample size, especially under
water background. 
In addition, we notice that nowadays most of the
experimental designs have allowed us to merge samples from
adjacent conditions (groups). At the end of the day we may not
have to rely on the alternative procedure for very small sample
size cases.   

Meanwhile, as the primary goal of this paper is to accurately
identify relevant biomarkers in discovery study, which is mostly balanced
design with moderate or small sample size, the method we
propose is mainly for the balanced design given two conditions. Thus how to
handle the data acquired from highly unbalanced design under
multiple conditions (more than two) with potentially very small
sample size for certain conditions is part of our future work. 

The nature of discovery study combined with the advancement of
DIA experiment result in sparse and high dimensional data. It is
known that popular univariate methods 
struggle under sparsity and multivariate methods stumble under high
dimensionality.  

Nonetheless, our proposed likelihood-based statistic resolves the
problem from a critical perspective -- accurately estimates the
number of truly relevant markers. This feature can effectively
resolves the aforesaid drawbacks from current methods.  

From multiple experimental and simulated data sets we demonstrate 
that our approach is able to accurately identify the set of truly
relevant protein markers. The implication is that our proposed method can
simplify the unnecessarily lengthy and costly process of biomarker discovery
without compromising the findings of markers correlating with
conditions.

\end{document}